\documentclass[%
aip,apl,
 amsmath,
 amssymb,
reprint,%
]{revtex4-1}

\usepackage{graphicx}
\usepackage{dcolumn}
\usepackage{bm}

\usepackage[utf8]{inputenc}
\usepackage[T1]{fontenc}
\usepackage{mathptmx}
\usepackage{siunitx}
\usepackage{subfigure}
\usepackage{textcomp} 
\usepackage{epstopdf}
\begin{document}

\preprint{AIP/123-QED}

\title{Atomically-Smooth Single-Crystalline VO$_2$ thin films with Bulk-like Metal-Insulator Transitions }

\author{Debasish Mondal}
\altaffiliation{These authors contributed equally}
\affiliation{Solid State and Structural Chemistry Unit, Indian institute of Science, Bengaluru, India}

\author{Smruti Rekha Mahapatra}
\altaffiliation{These authors contributed equally}
\affiliation{Solid State and Structural Chemistry Unit, Indian institute of Science, Bengaluru, India}

\author{Tanweer Ahmed}
\affiliation{Department of Physics, Indian institute of Science, Bengaluru, India}

\author{Suresh Kumar Podapangi}
\affiliation{Solid State and Structural Chemistry Unit, Indian institute of Science, Bengaluru, India}

\author{Arindam Ghosh}
\affiliation{Department of Physics, Indian institute of Science, Bengaluru, India}

\author{Naga Phani B. Aetukuri}
\email{phani@alumni.stanford.edu; phani@iisc.ac.in}
\affiliation{Solid State and Structural Chemistry Unit, Indian institute of Science, Bengaluru, India}

\date{\today}

\begin{abstract}
Atomically-abrupt interfaces in transition metal oxide (TMO) heterostructures could host a variety of exotic condensed matter phases that may not be found in the bulk materials at equilibrium. A critical step in the development of such atomically-sharp interfaces is the deposition of atomically-smooth TMO thin films. Optimized deposition conditions exist for the growth of perovskite oxides. However, the deposition of rutile oxides, such as VO$_2$, with atomic-layer precision has been challenging. In this work, we used pulsed laser deposition (PLD) to grow atomically-smooth VO$_2$ thin films on rutile TiO$_2$ (101) substrates. We show that optimal substrate preparation procedure followed by the deposition of VO$_2$ films at a temperature conducive for step-flow growth mode is essential for achieving atomically-smooth VO$_2$ films. The films deposited at optimal substrate temperatures show a step and terrace structure of the underlying TiO$_2$ substrate. At lower deposition temperatures, there is a transition to a mixed growth mode comprising of island growth and layer-by-layer growth modes. VO$_2$ films deposited at optimal substrate temperatures undergo a metal to insulator transition at a transition temperature of $\sim$325 K with $\sim$10$^3$ times increase in resistance, akin to MIT in bulk VO$_2$.
\end{abstract}
\maketitle
Growth of high quality single-crystalline thin films with atomically smooth surfaces is an essential prerequisite for the realization of heterostructures with atomically-abrupt interfaces. Such high quality interfaces could host exotic condensed matter phases that may otherwise not be realized in the bulk.\cite{kawasaki1994atomic,ohtomo2004high,reyren2007superconducting,brinkman2007magnetic} There is a large body of work that concerns the deposition and growth of perovskite oxide heterostructures with atomically sharp interfaces.\cite{mannhart2010oxide,thiel2006tunable,salluzzo2009orbital,chakhalian2007orbital} By contrast, research work on the growth of binary oxide heterostructures such as those comprising VO$_2$ and related rutile oxides, is limited.\cite{shibuya2010metal} This is rather surprising because the electronic properties of VO$_2$, a prototypical correlated electron material with a metal-insulator transition, have been extensively studied both in bulk and in thin film form.\cite{eyert2011vo,pardo2009half,qazilbash2007mott,haverkort2005orbital,berglund1969electronic,berglund1969hydrostatic} VO$_2$ undergoes a temperature-driven metal-insulator transition (MIT) from a high-temperature metallic phase to a low-temperature insulating phase at a characteristic phase transition temperature, T$_{MIT}$, of $\sim$340 K.\cite{goodenough1971two,morin1959oxides} This electronic transition is concomitant with a symmetry lowering structural transition from a high temperature rutile (P4$_2$/mnm) crystal structure to a low temperature monoclinic (P2$_1$/c) crystal structure. The manipulation of MIT in VO$_2$ via external perturbations such as strain, pressure, temperature and/or electric-field were previously attempted by several research groups in order to study the origin of MIT in this material.\cite{berglund1969hydrostatic,laverock2012strain,lazarovits2010effects,muraoka2002large,nakano2012collective,okazaki2004photoemission} In addition, the modification of the electronic structure of VO$_2$ in heterostructures is attractive for the creation of novel low power oxide electronic devices such as the Mott field effect transistor (Mott-FET).\cite{newns1998mott,yajima2015positive}

In this work, we show that single-crystalline thin films of VO$_2$ with near-ideal atomically smooth surfaces can be deposited on optimally prepared single-crystalline rutile TiO$_2$ substrates. We further show that the films retain bulk-like sharp transitions with $\sim$10$^3$ times changes in resistance across the transition. To the best of our knowledge, the realization of near-ideal surfaces and sharp metal-insulator transitions in the same set of films has not been achieved previously. For example, the growth of VO$_2$ on single-crystalline TiO$_2$ (001) substrates was shown to have sharp metal-insulator transitions.\cite{muraoka2002metal,aetukuri2013control,jeong2013suppression} Unfortunately, the metal-insulator transition is suppressed to $\sim$300 K or lower making these films not suitable for devices operating at room temperature. In addition, due to the high surface energy of the rutile (001) plane, annealing TiO$_2$ (001) substrates at temperatures that favor atomic step and terrace formation via surface recrystallization results in the undesired surface reconstruction and consequent surface roughening.\cite{ramamoorthy1994first,diebold2003surface,phanithesis} 

On the other hand, TiO$_2$ (110) surfaces have a low surface energy and are therefore amenable for annealing techniques to realize an atomic step and terrace structure.\cite{ramamoorthy1994first,diebold2003surface} However, the large and anisotropic in-plane strain imparted to the VO$_2$ films when epitaxially deposited on TiO$_2$ (110) substrates, usually leads to a broad metal-insulator transition.\cite{muraoka2002metal} Sharp metal-insulator transitions might lead to devices with high ON/OFF ratios and are therefore preferred for device applications. By contrast, when VO$_2$ films are grown on TiO$_2$ (101) surfaces, there is a smaller lattice mismatch when compared to the VO$_2$ films grown on TiO$_2$ (110) substrates. This could result in VO$_2$ films with a sharp MIT at a transition temperature that is closer to T$_{MIT}$ in bulk VO$_2$.\cite{li2014characteristics} Indeed, in agreement with this hypothesis, VO$_2$ films grown on TiO$_2$ (101) substrates have been shown to exhibit sharp transitions at T$_{MIT}$ of $\sim$320 K or greater.\cite{jeong2015giant} In addition, TiO$_2$ (101) surfaces have a surface energy that is intermediate between that of TiO$_2$ (110) and TiO$_2$ (001).\cite{ramamoorthy1994first} Therefore, it may be possible to achieve an atomic step and terrace structure, by annealing at an appropriate temperature favorable for surface recrystallization, while avoiding the undesired 3D surface reconstruction such as that observed on TiO$_2$ (001) surfaces.\cite{phanithesis} In the following sections, we outline a reproducible procedure for realizing a step and terrace structure on TiO$_2$ (101) substrates and show that PLD can be used to grow epitaxial VO$_2$ (101) films that retain the step and terrace structure of the underlying substrate while also showing sharp MIT at a T$_{MIT}$ of $\sim$325 K.

Single-crystalline rutile TiO$_2$ (101) substrates (Shinkosa, Japan) were treated using the following procedure. First, to remove any organic layers on the surface, as-received substrates were ultrasonically cleaned in an acetone bath (ACS Reagent grade with 99.5\% purity from Merck) for 15 minutes followed by ultrasonic cleaning in an isopropyl alcohol (IPA) bath (ACS Reagent grade with 99.5\% purity from Merck) for 5 minutes. Substrates were then rinsed in deionized water (resistivity 17.5 M$\Omega$-cm) and blow dried with nitrogen. Next, to remove trace metal contaminants from the surface, the substrates were placed in a still solution of 5 vol. \% hydrochloric acid (HCl) for 5 minutes followed by rinsing with DI water and nitrogen blow drying. The substrates were then transferred to a still 50\% hydrofluoric acid (HF) bath.
\begin{figure}[!htb]
	\includegraphics[width=0.47\textwidth]{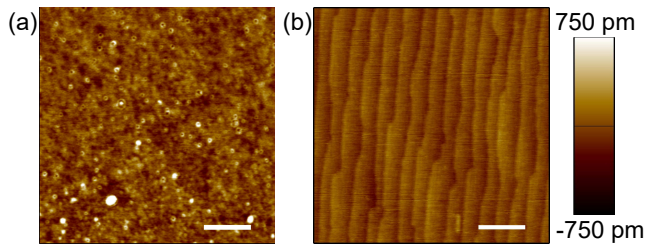}
	\centering
	\caption{AFM images of TiO$_2$ (101) substrates (a) after the chemical cleaning procedure but before annealing and (b) after annealing at 1000 °C for 4 hours under an oxygen flow rate of 10 sccm. Atomic step and terrace structure is realized after the annealing step. The scale bars are equivalent to 400 nm. The images are taken in tapping mode (AC Air topography) using an Asylum Cypher ES AFM.}
\end{figure}
The substrates were taken out after 1 minute and immediately rinsed with DI water and blow dried with nitrogen. This chemical cleaning procedure is similar to the one previously reported by Martens et al.,\cite{martens2014improved} for cleaning TiO$_2$ (001) surfaces. 

After the chemical cleaning steps (Figure 1a), the substrates were annealed at temperatures between 950 °C and 1000 °C for 4 hours under an oxygen flow rate of 10 sccm. The rate of warm up and cool down were both set to 3 °C/min. Based on our experiments, it seems that a temperature of $\sim$950 °C is necessary for recrystallization leading to the formation of a step and terrace structure (Figure 1b) with each step height being equivalent to the inter planner spacing of (101) planes. The step and terrace structure is the result of a small but non-zero miscut angle between the substrate surface and the surface of the TiO$_2$ (101) crystallographic plane. Such surfaces are commonly referred to as vicinal surfaces. All substrates used for the growth of VO$_2$ films in this work were annealed at temperatures between 950 °C and 1000 °C, as step-bunching was observed at temperatures higher than 1000 °C.
\begin{figure}[!htb]
	\setlength\belowcaptionskip{-2.6ex}
	\centering
	\begin{subfigure}
		\centering
		\includegraphics[width=0.47\textwidth]{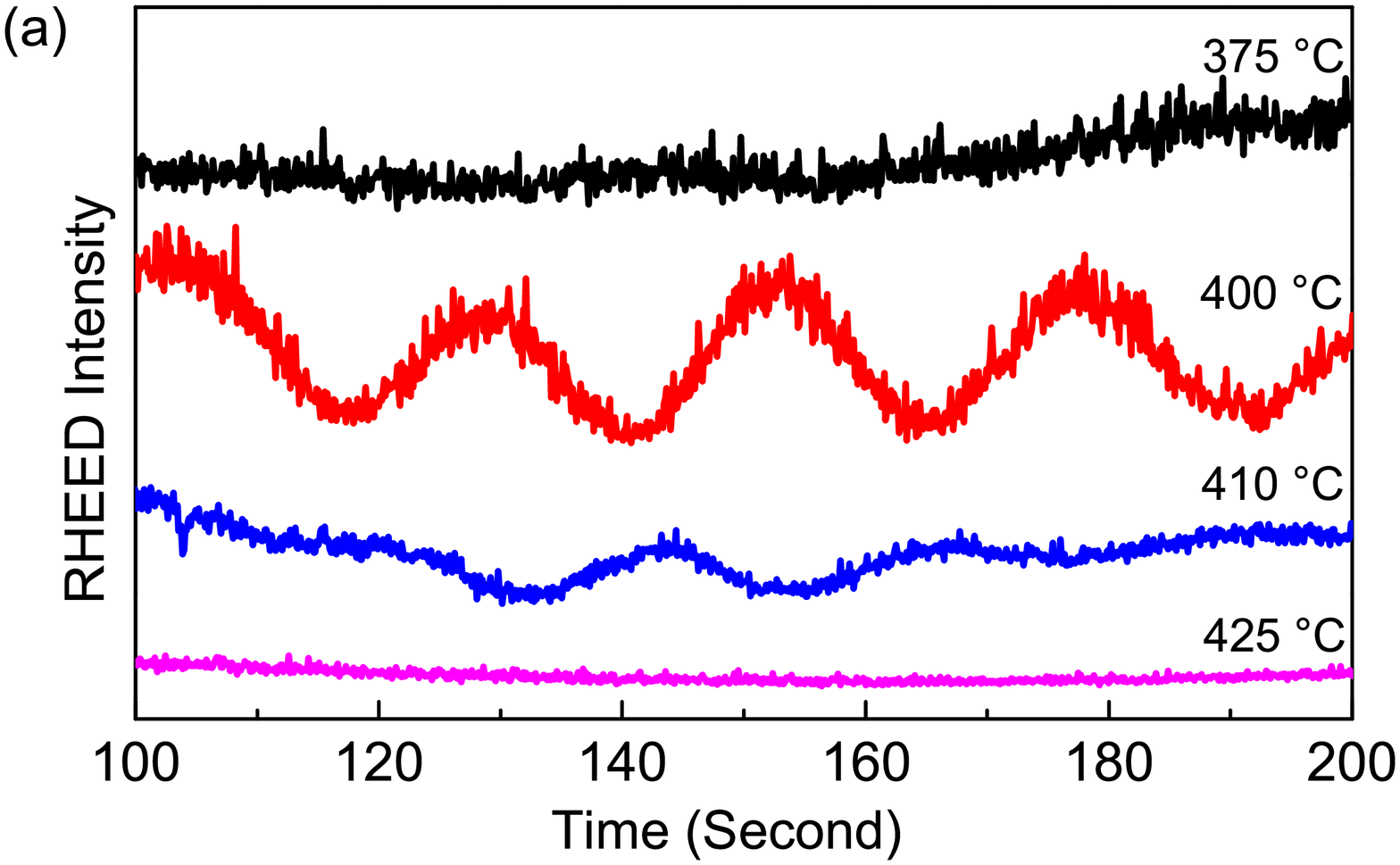}
	\end{subfigure}
	\begin{subfigure}
		\centering
		\includegraphics[width=0.47\textwidth]{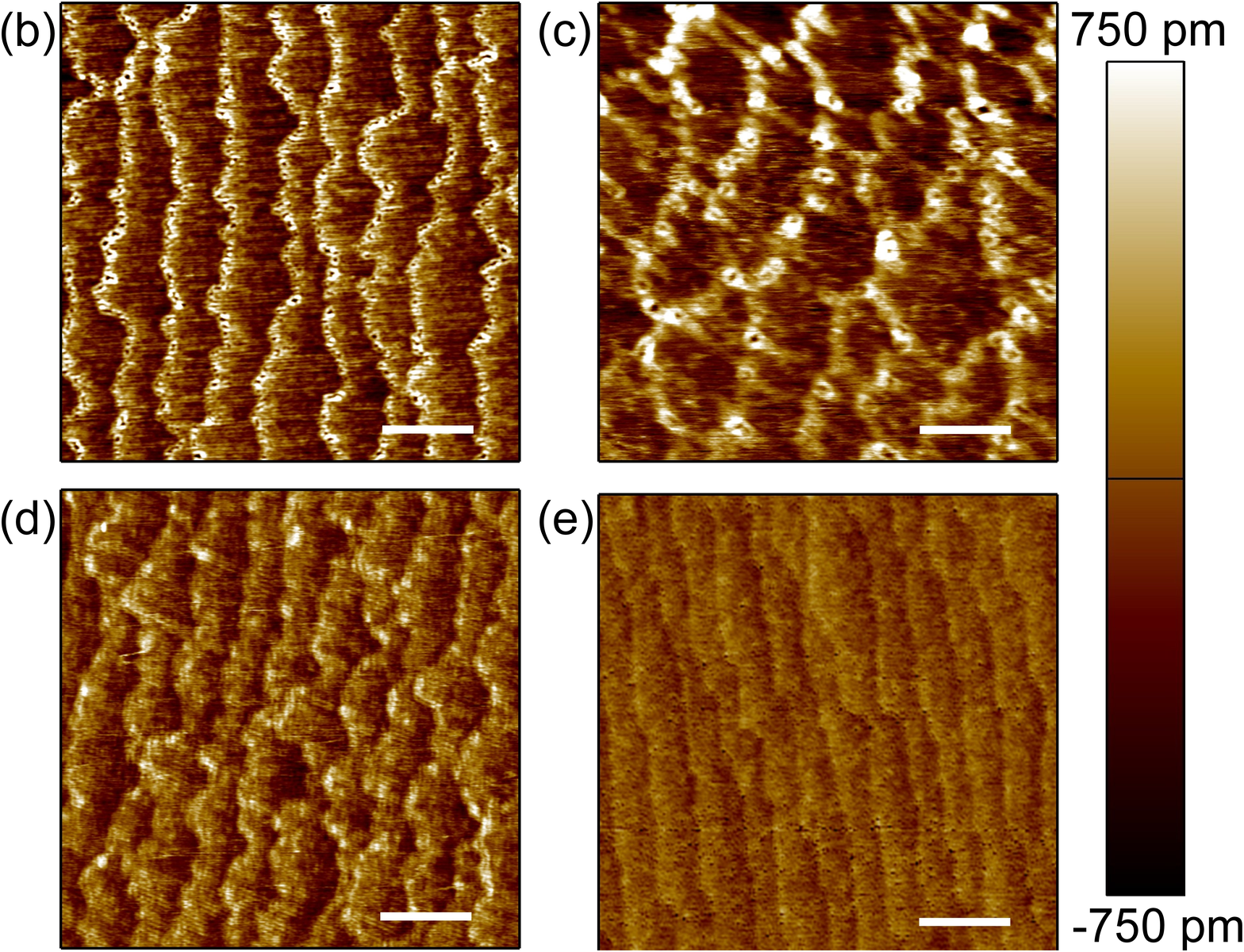}
	\end{subfigure}
	\begin{subfigure}
		\centering
		\includegraphics[width=0.47\textwidth]{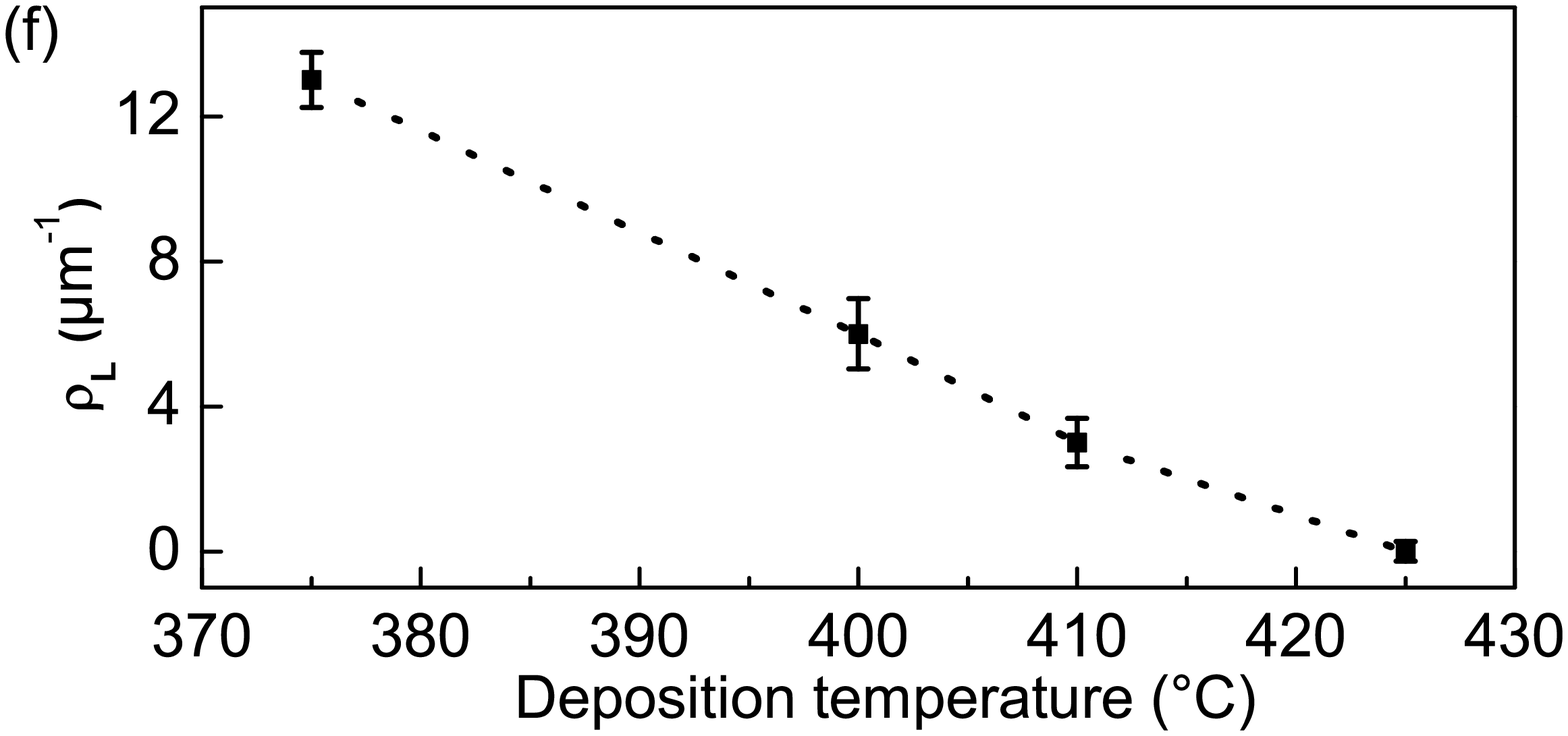}
	\end{subfigure}
	\caption{(a) RHEED intensity oscillations during VO$_2$ deposition at different substrate temperatures as mentioned in the figure. Here the curves are offset for clarity. AFM topographic images of 4 nm VO$_2$ thin films deposited on TiO$_2$ (101) substrates at substrate temperatures of (b) 375 °C, (c) 400 °C, (d) 410 °C and (e) 425 °C. Atomically smooth step and terrace structure was obtained at a temperature of 425 °C. At substrate temperatures below 425 °C, a combination of 3D islands at step edges and atomically smooth terraces were observed. The scale bars are equivalent to 400 nm in all AFM images. (f) A plot of linear density of islands ($\rho_L$) versus the substrate temperature showing a monotonic decrease in the number of islands at the step edges as the substrate temperature during deposition is increased.}
\end{figure}

On vicinal surfaces, such as the surfaces of TiO$_2$ substrates used in this work, with a near ideal step and terrace structure, heteroepitaxial growth could proceed by one or more of the three broadly classified growth modes: a) three dimensional (3D) island-growth mode (also called Volmer-Weber growth mode); b) a mixture of 2-dimensional (2D) layer-by-layer growth and 3D island growth modes (Stranski-Krastanov growth mode) and c) 2D growth mode: either layer-by-layer growth or step-flow growth modes.\cite{gilmer1987models,hong2005persistent} The preferred growth mode(s) is highly sensitive to the surface diffusivity of the constituent species (individual growth units) on the substrate surface. For example, substrate temperature is a simple yet effective parameter to tune the surface diffusivity of the growth units and therefore the growth mode.\cite{hong2005persistent,koster1999imposed} A near-ideal atomic step and terrace structure is realized when layer-by-layer and/or step-flow growth modes are the dominant growth modes. Therefore, in this work, we varied the substrate temperature while the other growth conditions were not changed as discussed below. 

We employed PLD for the growth of VO$_2$ thin films on optimally annealed rutile TiO$_2$ (101) substrates (5 mm $\times$ 5 mm $\times$ 0.5 mm). All depositions were performed in a chamber with a base pressure of <5 $\times$ 10$^{-9}$ Torr. A V$_2$O$_5$ pellet, prepared by compacting V$_2$O$_5$ powder ($\geq$99.6\% trace metals basis from Merck) followed by sintering at 670 °C in air for 10 hours, was used as the target. Laser energy density on the target during ablation of $\sim$1.5 J/cm$^2$, laser repetition rate of 2 Hz, substrate to target distance of $\sim$55 mm and a background deposition oxygen pressure of 10 mTorr were used for all depositions reported in this work. Under these conditions, the growth rate was calculated (based on thickness measurements by a profilometer) to be $\sim$0.09 \AA/s. As discussed later, this growth rate is also validated by in-situ high pressure reflection high energy electron diffraction (hp-RHEED) which was used to monitor growth and surface structure during deposition.

The results of temperature-dependent growth dynamics for 4 nm VO$_2$ thin films grown on optimally annealed TiO$_2$ (101) substrates are summarized in Figure 2. At the lowest temperature of  375 °C, we did not observe any RHEED intensity oscillations (Figure 2a); corresponding atomic force microscopy image, Figure 2b, shows an atomic step and terrace structure similar to the surface structure of the starting substrate. Notably, upon closer inspection, we observed islands at step edges suggesting that the growth mode at this temperature is a mixture of 3D island growth (at step edges) and layer-by-layer growth on the steps. At a higher substrate temperature of 400 °C, as shown in Figure 2a, we observed RHEED intensity oscillations. The average growth rate of $\sim$0.09 \AA/s estimated from RHEED intensity oscillations is consistent with the growth rates estimated from profilometer measurements. However, RHEED intensity oscillations are not sufficient evidence for a truly 2-dimensional (2D) layer-by-layer growth. In fact, AFM measurements on the VO$_2$ thin film surfaces grown at 400 °C (Figure 2c), show 3D islands at the step edges similar to the observation on the VO$_2$ thin films deposited at a substrate temperature of 375 °C. Clearly, the linear density of islands, defined as the number of islands per unit length along the step edge and plotted in Figure 2f, decreased at the higher deposition temperature of 400 °C. We take this as indirect evidence for increased surface diffusivity of the growth units at the higher temperature. At an even higher substrate temperature of 410 °C, there is a further decrease in the linear density of islands as evidenced from AFM image shown in Figure 2d. Furthermore, the amplitude of the RHEED intensity oscillations also decreases. 
\begin{figure}[!htb]
	\setlength\belowcaptionskip{-2.7ex}
	\centering
	\begin{subfigure}
		\centering
		\includegraphics[width=0.47\textwidth]{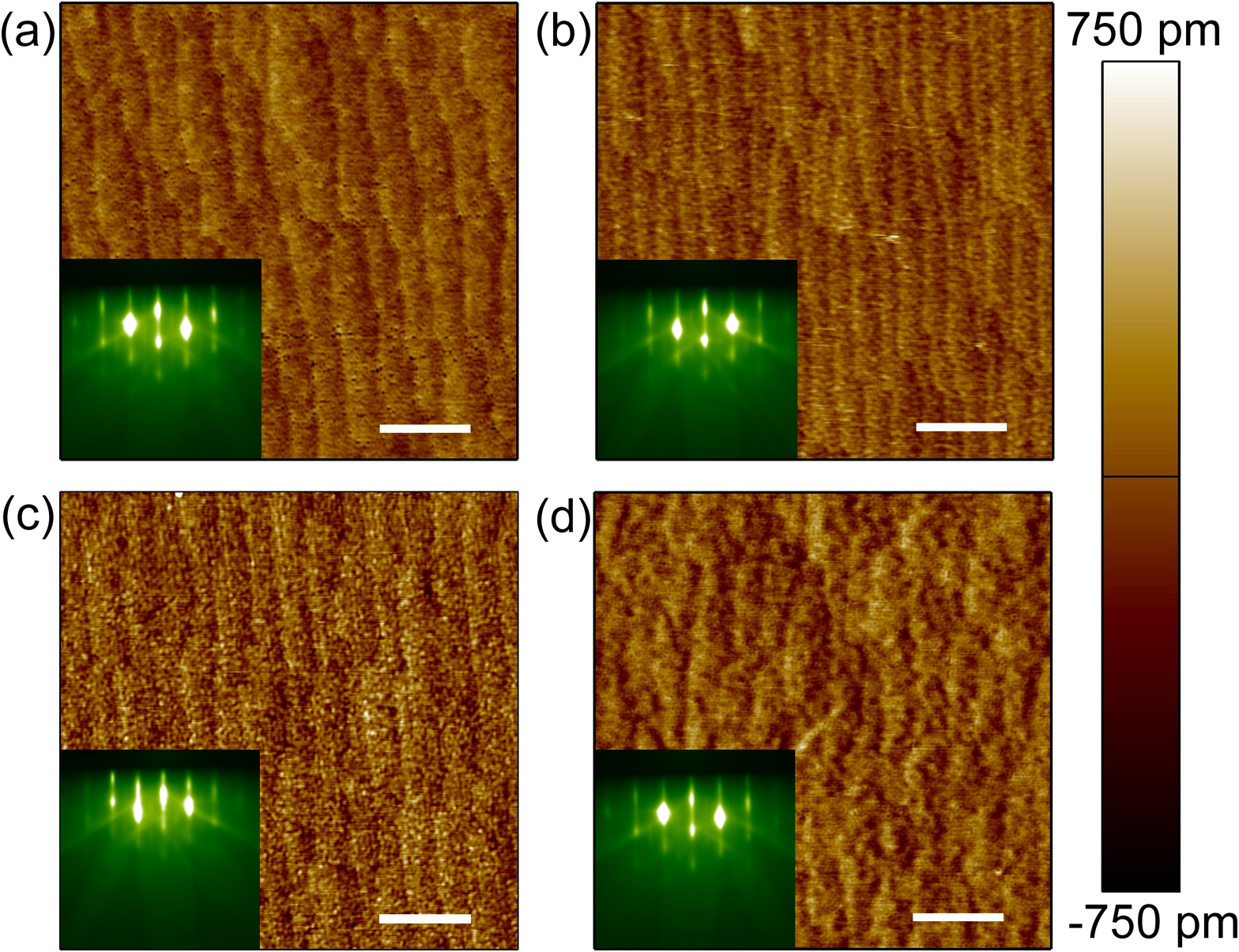}
	\end{subfigure}
	\begin{subfigure}
		\centering
		\includegraphics[width=0.47\textwidth]{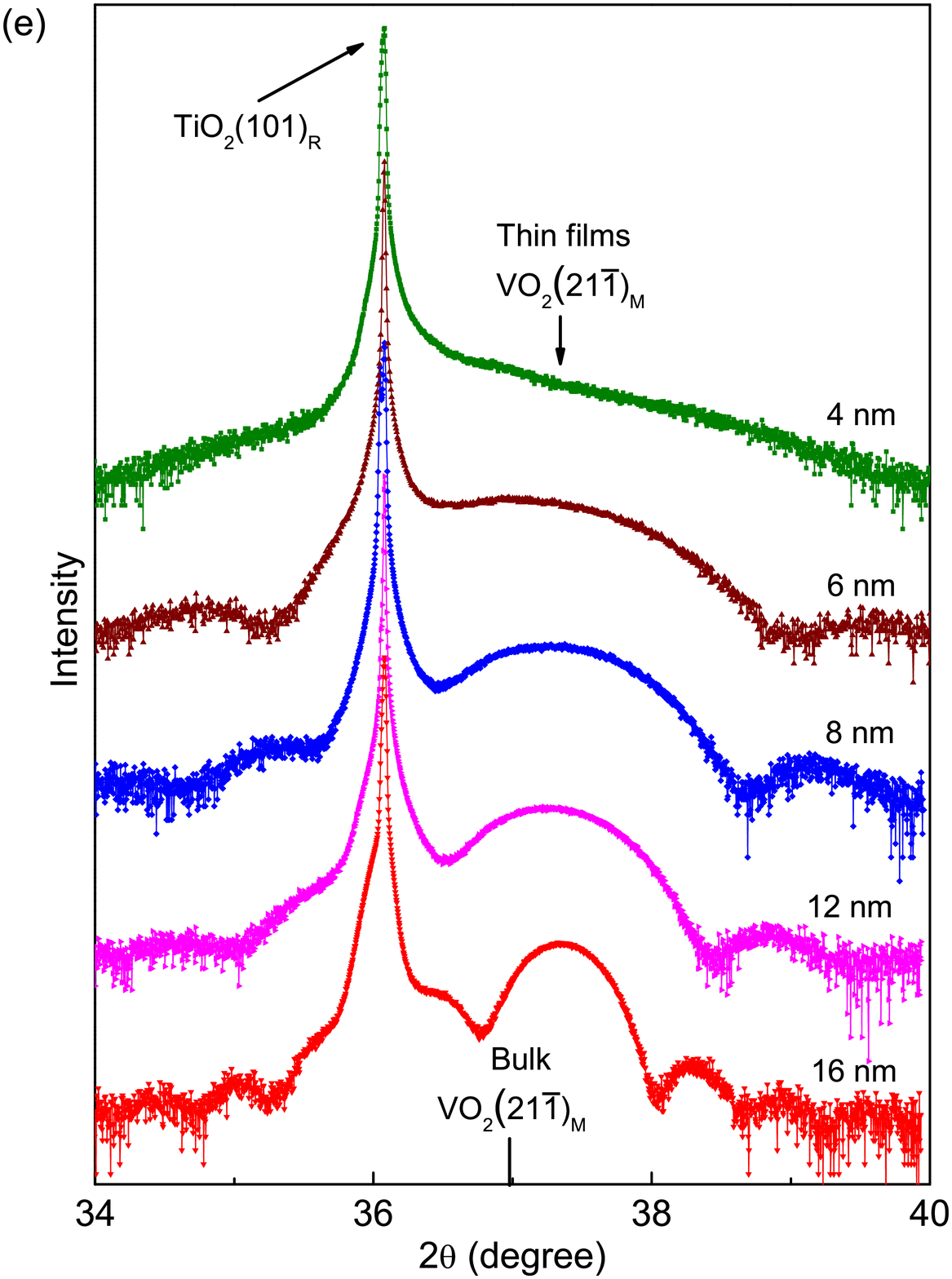}
	\end{subfigure}
	\caption{AFM images of VO$_2$ thin films deposited at 425 °C but with different thicknesses of (a) 4 nm, (b) 8 nm, (c) 12 nm and (d) 16 nm. The insets show the corresponding hp-RHEED patterns of the films along the crystallographic [${\bar{1}01}$] direction. The observed RHEED patterns are indicative of smooth and highly crystalline VO$_2$ thin films. (e) Typical high resolution $\theta$-2$\theta$ XRD scans of VO$_2$ thin films of different thicknesses ranging from 4 nm to 16 nm as mentioned in the figure legend. XRD data is indicative of coherently strained and highly oriented or single-crystalline VO$_2$ thin films on TiO$_2$ (101) substrates. The expected position of the bulk VO$_2$ (${21\bar{1}}$)$_M$ peak (solid vertical bar at $\sim$\ang{37}) and the approximate position of VO$_2$ (${21\bar{1}}$)$_M$  peak for the thin films (solid vertical arrow at $\sim$\ang{37.4}) are also shown. The curves in Figure 3e are offset for clarity.}
\end{figure}

Finally, as shown in Figure 2e, at the highest substrate temperature of 425 °C used in this study, we did not observe any RHEED intensity oscillations. Furthermore, the surface of the thin films showed the atomic step and terrace structure of the underlying substrate without any discernible islands at the step edges. The absence of RHEED intensity oscillations together with the observation of an atomic step and terrace structure is indicative of 2D step flow growth mode. Clearly, these changes in topography and the amplitude of RHEED intensity oscillations are consistent with the growth mode changing from a growth mode comprising of a mixture of 3D island growth and 2D layer-by-layer growth at the substrate temperature of 375 °C to a 2D step-flow growth mode at the substrate temperature of 425 °C. To the best of our knowledge, such a difference in the growth kinetics at a step edge and on the atomic steps has not been reported before. An interesting future research direction could be to explore if this difference can be exploited for directed 3D growth of, for example, VO$_2$ nanorods on TiO$_2$ substrates. Such directional 3D structures might find applications in tunable metamaterials based on metal-insulator transitions.\cite{liu2012terahertz,dicken2009frequency,kabashin2009plasmonic}

Next, we show that this step and terrace structure can be retained for VO$_2$ film thicknesses of up to 16 nm, when thin films are deposited at the optimal substrate temperature of 425 °C. AFM images shown in Figures 3a-d for film thickness of 4 nm to 16 nm are indicative of the underlying step and terrace structure of the TiO$_2$ substrate being retained. Furthermore, RHEED streaks that are evident in the RHEED diffraction patterns, shown in the insets of the corresponding AFM images, are suggestive of an atomically smooth 2D surface.\cite{Ichimiya2004RHEED} This is also evidence for the single-crystalline nature of the thin films. At much higher thicknesses, for example at a film thickness of 32 nm, we have observed cracking of the VO$_2$ films: possibly due to interfacial strain.\cite{kawatani2014formation} Therefore, the rest of our studies are limited to films with thicknesses of $\leq$16 nm. 

High resolution $\theta$-2$\theta$ X-ray diffraction (XRD) measurements (Rigaku SmartLab X-ray diffractometer) performed on VO$_2$ thin films with thicknesses of 4 nm, 6 nm, 8 nm, 12 nm and 16 nm  are evidence for the excellent crystallinity of the VO$_2$ films (Figure 3e). 
\begin{figure}[!htb]
	\setlength\belowcaptionskip{-2.5ex}
	\includegraphics[width=0.47\textwidth]{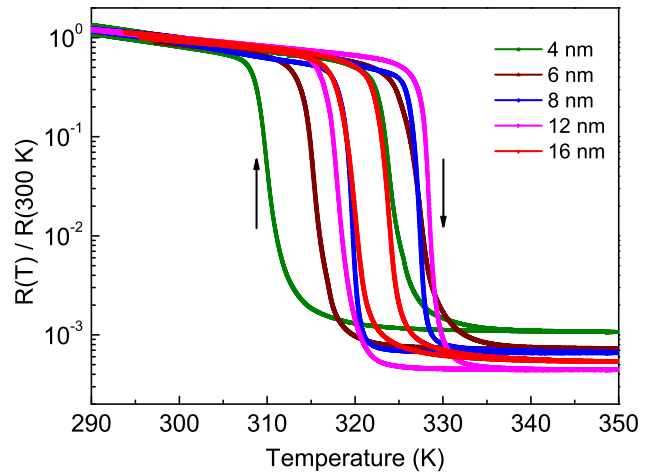}
	\centering
	\caption{Resistance versus temperature plots of VO$_2$ thin films with the different thicknesses. At 300 K, resistance of 4 nm, 6 nm, 8 nm, 12 nm and 16 nm films are 439 k$\Omega$, 253 k$\Omega$, 145 k$\Omega$, 238 k$\Omega$ and 146 k$\Omega$ respectively. All films have a nearly identical T$_{MIT}$ of $\sim$325 K and a nearly 3 orders of change in resistance across the metal-insulator transition. The resistance is plotted normalized to the resistance at 300 K.}
\end{figure} 
The increase in peak intensity and the decrease of full width at half maximum of the VO$_2$ (21$\bar1$)$_M$ peak is in agreement with the increase in the thickness of VO$_2$ films. Additionally, the presence of Keissig fringes in the diffractogram is indicative of a sharp interface between the VO$_2$ film and the TiO$_2$ substrate. 

A summary of the resistance versus temperature measurements (four probe geometry using Keithley 2400 source measure units) performed while continuously heating or cooling the sample at a rate of 1 K/min is shown in Figure 4. All VO$_2$ films show sharp metal-insulator transitions at an average T$_{MIT}$ of >320 K with $\sim$10$^3$ times change in resistance. Resistance normalized to the resistance at 300 K is plotted to compare the magnitude of the change in resistance of all the films; evidently, this is nearly identical for all thicknesses of VO$_2$ films studied in this work. In addition, we did not observe any systematic changes in the transition temperature as a function of the thickness of VO$_2$ films. There seems to be a uniform reduction in the T$_{MIT}$ for all films, possibly due to the VO$_2$ film being epitaxially strained by the TiO$_2$ substrate. This is also consistent with the X-ray measurements which show that all films in the thickness range of 4 nm to 16 nm are epitaxially strained to the substrate. An out-of-plane compressive strain of $\sim$0.9\% was inferred from XRD measurements (see Figure 3e and supplementary information).

In summary, our work outlines a procedure to realize an atomic step and terrace structure on vicinal surfaces of rutile TiO$_2$ (101) substrates. We also demonstrate that a near-ideal atomic step and terrace structure can be realized by depositing VO$_2$ films at temperatures conducive for step flow growth. We have also observed a previously not reported mixed growth mode that comprises of a 3D growth mode at step edges and a 2D layer-by-layer growth mode on the atomic steps. Importantly, our results imply that VO$_2$ (101) thin films as thin as 4 nm can have bulk-like transitions at a T$_{MIT}$ well above room temperature with $\Delta$R/R of $\sim$10$^3$ and are therefore suitable for device applications. Furthermore, under optimal conditions, we showed that the near ideal atomic step and terrace structure can be stabilized in VO$_2$ films with thicknesses upto 16 nm. The growth of such atomically smooth VO$_2$ thin films with bulk-like transitions and heterostructures comprising other related oxides such as CrO$_2$\cite{cai2015single} could open up the possibility of developing binary oxide heterostructures. Such heterostructures could enable interfacial manipulation of metal-insulator transitions and/or play host to exotic condensed matter phases not realizable in the bulk.

\vspace{10pt}
The authors acknowledge CeNSE, IISc for access to HR-XRD, wire bonding and clean-room facilities. N.B.A acknowledges the new faculty start-up grant no. 12-0205-0618-77 provided by the Indian Institute of Science. N.B.A is thankful to Prof. D D Sarma and Prof. Anil Kumar for access to the PLD system. S.R.M and D.M want to thank Jibin J. Samuel for useful discussions. We thank Prof. Satish Patil for providing access to facilities supported by the Swarnajayanti fellowship under grant No. DST/SJF/CSA-01/2013-14.
\appendix
\nocite{*}
\bibliography{referanceVO2Work}

\begin{thebibliography}{42}%
\makeatletter
\providecommand \@ifxundefined [1]{%
 \@ifx{#1\undefined}
}%
\providecommand \@ifnum [1]{%
 \ifnum #1\expandafter \@firstoftwo
 \else \expandafter \@secondoftwo
 \fi
}%
\providecommand \@ifx [1]{%
 \ifx #1\expandafter \@firstoftwo
 \else \expandafter \@secondoftwo
 \fi
}%
\providecommand \natexlab [1]{#1}%
\providecommand \enquote  [1]{``#1''}%
\providecommand \bibnamefont  [1]{#1}%
\providecommand \bibfnamefont [1]{#1}%
\providecommand \citenamefont [1]{#1}%
\providecommand \href@noop [0]{\@secondoftwo}%
\providecommand \href [0]{\begingroup \@sanitize@url \@href}%
\providecommand \@href[1]{\@@startlink{#1}\@@href}%
\providecommand \@@href[1]{\endgroup#1\@@endlink}%
\providecommand \@sanitize@url [0]{\catcode `\\12\catcode `\$12\catcode
  `\&12\catcode `\#12\catcode `\^12\catcode `\_12\catcode `\%12\relax}%
\providecommand \@@startlink[1]{}%
\providecommand \@@endlink[0]{}%
\providecommand \url  [0]{\begingroup\@sanitize@url \@url }%
\providecommand \@url [1]{\endgroup\@href {#1}{\urlprefix }}%
\providecommand \urlprefix  [0]{URL }%
\providecommand \Eprint [0]{\href }%
\providecommand \doibase [0]{http://dx.doi.org/}%
\providecommand \selectlanguage [0]{\@gobble}%
\providecommand \bibinfo  [0]{\@secondoftwo}%
\providecommand \bibfield  [0]{\@secondoftwo}%
\providecommand \translation [1]{[#1]}%
\providecommand \BibitemOpen [0]{}%
\providecommand \bibitemStop [0]{}%
\providecommand \bibitemNoStop [0]{.\EOS\space}%
\providecommand \EOS [0]{\spacefactor3000\relax}%
\providecommand \BibitemShut  [1]{\csname bibitem#1\endcsname}%
\let\auto@bib@innerbib\@empty
\bibitem [{\citenamefont {Kawasaki}\ \emph {et~al.}(1994)\citenamefont
  {Kawasaki}, \citenamefont {Takahashi}, \citenamefont {Maeda}, \citenamefont
  {Tsuchiya}, \citenamefont {Shinohara}, \citenamefont {Ishiyama},
  \citenamefont {Yonezawa}, \citenamefont {Yoshimoto},\ and\ \citenamefont
  {Koinuma}}]{kawasaki1994atomic}%
  \BibitemOpen
  \bibfield  {author} {\bibinfo {author} {\bibfnamefont {M.}~\bibnamefont
  {Kawasaki}}, \bibinfo {author} {\bibfnamefont {K.}~\bibnamefont {Takahashi}},
  \bibinfo {author} {\bibfnamefont {T.}~\bibnamefont {Maeda}}, \bibinfo
  {author} {\bibfnamefont {R.}~\bibnamefont {Tsuchiya}}, \bibinfo {author}
  {\bibfnamefont {M.}~\bibnamefont {Shinohara}}, \bibinfo {author}
  {\bibfnamefont {O.}~\bibnamefont {Ishiyama}}, \bibinfo {author}
  {\bibfnamefont {T.}~\bibnamefont {Yonezawa}}, \bibinfo {author}
  {\bibfnamefont {M.}~\bibnamefont {Yoshimoto}}, \ and\ \bibinfo {author}
  {\bibfnamefont {H.}~\bibnamefont {Koinuma}},\ }\href@noop {} {\bibfield
  {journal} {\bibinfo  {journal} {Science}\ }\textbf {\bibinfo {volume}
  {266}},\ \bibinfo {pages} {1540} (\bibinfo {year} {1994})}\BibitemShut
  {NoStop}%
\bibitem [{\citenamefont {Ohtomo}\ and\ \citenamefont
  {Hwang}(2004)}]{ohtomo2004high}%
  \BibitemOpen
  \bibfield  {author} {\bibinfo {author} {\bibfnamefont {A.}~\bibnamefont
  {Ohtomo}}\ and\ \bibinfo {author} {\bibfnamefont {H.}~\bibnamefont {Hwang}},\
  }\href@noop {} {\bibfield  {journal} {\bibinfo  {journal} {Nature}\ }\textbf
  {\bibinfo {volume} {427}},\ \bibinfo {pages} {423} (\bibinfo {year}
  {2004})}\BibitemShut {NoStop}%
\bibitem [{\citenamefont {Reyren}\ \emph {et~al.}(2007)\citenamefont {Reyren},
  \citenamefont {Thiel}, \citenamefont {Caviglia}, \citenamefont {Kourkoutis},
  \citenamefont {Hammerl}, \citenamefont {Richter}, \citenamefont {Schneider},
  \citenamefont {Kopp}, \citenamefont {R{\"u}etschi}, \citenamefont {Jaccard}
  \emph {et~al.}}]{reyren2007superconducting}%
  \BibitemOpen
  \bibfield  {author} {\bibinfo {author} {\bibfnamefont {N.}~\bibnamefont
  {Reyren}}, \bibinfo {author} {\bibfnamefont {S.}~\bibnamefont {Thiel}},
  \bibinfo {author} {\bibfnamefont {A.}~\bibnamefont {Caviglia}}, \bibinfo
  {author} {\bibfnamefont {L.~F.}\ \bibnamefont {Kourkoutis}}, \bibinfo
  {author} {\bibfnamefont {G.}~\bibnamefont {Hammerl}}, \bibinfo {author}
  {\bibfnamefont {C.}~\bibnamefont {Richter}}, \bibinfo {author} {\bibfnamefont
  {C.}~\bibnamefont {Schneider}}, \bibinfo {author} {\bibfnamefont
  {T.}~\bibnamefont {Kopp}}, \bibinfo {author} {\bibfnamefont {A.-S.}\
  \bibnamefont {R{\"u}etschi}}, \bibinfo {author} {\bibfnamefont
  {D.}~\bibnamefont {Jaccard}},  \emph {et~al.},\ }\href@noop {} {\bibfield
  {journal} {\bibinfo  {journal} {Science}\ }\textbf {\bibinfo {volume}
  {317}},\ \bibinfo {pages} {1196} (\bibinfo {year} {2007})}\BibitemShut
  {NoStop}%
\bibitem [{\citenamefont {Brinkman}\ \emph {et~al.}(2007)\citenamefont
  {Brinkman}, \citenamefont {Huijben}, \citenamefont {Van~Zalk}, \citenamefont
  {Huijben}, \citenamefont {Zeitler}, \citenamefont {Maan}, \citenamefont
  {van~der Wiel}, \citenamefont {Rijnders}, \citenamefont {Blank},\ and\
  \citenamefont {Hilgenkamp}}]{brinkman2007magnetic}%
  \BibitemOpen
  \bibfield  {author} {\bibinfo {author} {\bibfnamefont {A.}~\bibnamefont
  {Brinkman}}, \bibinfo {author} {\bibfnamefont {M.}~\bibnamefont {Huijben}},
  \bibinfo {author} {\bibfnamefont {M.}~\bibnamefont {Van~Zalk}}, \bibinfo
  {author} {\bibfnamefont {J.}~\bibnamefont {Huijben}}, \bibinfo {author}
  {\bibfnamefont {U.}~\bibnamefont {Zeitler}}, \bibinfo {author} {\bibfnamefont
  {J.}~\bibnamefont {Maan}}, \bibinfo {author} {\bibfnamefont {W.~G.}\
  \bibnamefont {van~der Wiel}}, \bibinfo {author} {\bibfnamefont
  {G.}~\bibnamefont {Rijnders}}, \bibinfo {author} {\bibfnamefont {D.~H.}\
  \bibnamefont {Blank}}, \ and\ \bibinfo {author} {\bibfnamefont
  {H.}~\bibnamefont {Hilgenkamp}},\ }\href@noop {} {\bibfield  {journal}
  {\bibinfo  {journal} {Nat. Mater.}\ }\textbf {\bibinfo {volume} {6}},\
  \bibinfo {pages} {493} (\bibinfo {year} {2007})}\BibitemShut {NoStop}%
\bibitem [{\citenamefont {Mannhart}\ and\ \citenamefont
  {Schlom}(2010)}]{mannhart2010oxide}%
  \BibitemOpen
  \bibfield  {author} {\bibinfo {author} {\bibfnamefont {J.}~\bibnamefont
  {Mannhart}}\ and\ \bibinfo {author} {\bibfnamefont {D.}~\bibnamefont
  {Schlom}},\ }\href@noop {} {\bibfield  {journal} {\bibinfo  {journal}
  {Science}\ }\textbf {\bibinfo {volume} {327}},\ \bibinfo {pages} {1607}
  (\bibinfo {year} {2010})}\BibitemShut {NoStop}%
\bibitem [{\citenamefont {Thiel}\ \emph {et~al.}(2006)\citenamefont {Thiel},
  \citenamefont {Hammerl}, \citenamefont {Schmehl}, \citenamefont {Schneider},\
  and\ \citenamefont {Mannhart}}]{thiel2006tunable}%
  \BibitemOpen
  \bibfield  {author} {\bibinfo {author} {\bibfnamefont {S.}~\bibnamefont
  {Thiel}}, \bibinfo {author} {\bibfnamefont {G.}~\bibnamefont {Hammerl}},
  \bibinfo {author} {\bibfnamefont {A.}~\bibnamefont {Schmehl}}, \bibinfo
  {author} {\bibfnamefont {C.}~\bibnamefont {Schneider}}, \ and\ \bibinfo
  {author} {\bibfnamefont {J.}~\bibnamefont {Mannhart}},\ }\href@noop {}
  {\bibfield  {journal} {\bibinfo  {journal} {Science}\ }\textbf {\bibinfo
  {volume} {313}},\ \bibinfo {pages} {1942} (\bibinfo {year}
  {2006})}\BibitemShut {NoStop}%
\bibitem [{\citenamefont {Salluzzo}\ \emph {et~al.}(2009)\citenamefont
  {Salluzzo}, \citenamefont {Cezar}, \citenamefont {Brookes}, \citenamefont
  {Bisogni}, \citenamefont {De~Luca}, \citenamefont {Richter}, \citenamefont
  {Thiel}, \citenamefont {Mannhart}, \citenamefont {Huijben}, \citenamefont
  {Brinkman} \emph {et~al.}}]{salluzzo2009orbital}%
  \BibitemOpen
  \bibfield  {author} {\bibinfo {author} {\bibfnamefont {M.}~\bibnamefont
  {Salluzzo}}, \bibinfo {author} {\bibfnamefont {J.}~\bibnamefont {Cezar}},
  \bibinfo {author} {\bibfnamefont {N.}~\bibnamefont {Brookes}}, \bibinfo
  {author} {\bibfnamefont {V.}~\bibnamefont {Bisogni}}, \bibinfo {author}
  {\bibfnamefont {G.}~\bibnamefont {De~Luca}}, \bibinfo {author} {\bibfnamefont
  {C.}~\bibnamefont {Richter}}, \bibinfo {author} {\bibfnamefont
  {S.}~\bibnamefont {Thiel}}, \bibinfo {author} {\bibfnamefont
  {J.}~\bibnamefont {Mannhart}}, \bibinfo {author} {\bibfnamefont
  {M.}~\bibnamefont {Huijben}}, \bibinfo {author} {\bibfnamefont
  {A.}~\bibnamefont {Brinkman}},  \emph {et~al.},\ }\href@noop {} {\bibfield
  {journal} {\bibinfo  {journal} {Phys. rev. lett.}\ }\textbf {\bibinfo
  {volume} {102}},\ \bibinfo {pages} {166804} (\bibinfo {year}
  {2009})}\BibitemShut {NoStop}%
\bibitem [{\citenamefont {Chakhalian}\ \emph {et~al.}(2007)\citenamefont
  {Chakhalian}, \citenamefont {Freeland}, \citenamefont {Habermeier},
  \citenamefont {Cristiani}, \citenamefont {Khaliullin}, \citenamefont
  {Van~Veenendaal},\ and\ \citenamefont {Keimer}}]{chakhalian2007orbital}%
  \BibitemOpen
  \bibfield  {author} {\bibinfo {author} {\bibfnamefont {J.}~\bibnamefont
  {Chakhalian}}, \bibinfo {author} {\bibfnamefont {J.}~\bibnamefont
  {Freeland}}, \bibinfo {author} {\bibfnamefont {H.-U.}\ \bibnamefont
  {Habermeier}}, \bibinfo {author} {\bibfnamefont {G.}~\bibnamefont
  {Cristiani}}, \bibinfo {author} {\bibfnamefont {G.}~\bibnamefont
  {Khaliullin}}, \bibinfo {author} {\bibfnamefont {M.}~\bibnamefont
  {Van~Veenendaal}}, \ and\ \bibinfo {author} {\bibfnamefont {B.}~\bibnamefont
  {Keimer}},\ }\href@noop {} {\bibfield  {journal} {\bibinfo  {journal}
  {Science}\ }\textbf {\bibinfo {volume} {318}},\ \bibinfo {pages} {1114}
  (\bibinfo {year} {2007})}\BibitemShut {NoStop}%
\bibitem [{\citenamefont {Shibuya}, \citenamefont {Kawasaki},\ and\
  \citenamefont {Tokura}(2010)}]{shibuya2010metal}%
  \BibitemOpen
  \bibfield  {author} {\bibinfo {author} {\bibfnamefont {K.}~\bibnamefont
  {Shibuya}}, \bibinfo {author} {\bibfnamefont {M.}~\bibnamefont {Kawasaki}}, \
  and\ \bibinfo {author} {\bibfnamefont {Y.}~\bibnamefont {Tokura}},\
  }\href@noop {} {\bibfield  {journal} {\bibinfo  {journal} {Phys. Rev. B}\
  }\textbf {\bibinfo {volume} {82}},\ \bibinfo {pages} {205118} (\bibinfo
  {year} {2010})}\BibitemShut {NoStop}%
\bibitem [{\citenamefont {Eyert}(2011)}]{eyert2011vo}%
  \BibitemOpen
  \bibfield  {author} {\bibinfo {author} {\bibfnamefont {V.}~\bibnamefont
  {Eyert}},\ }\href@noop {} {\bibfield  {journal} {\bibinfo  {journal} {Phys.
  Rev. Lett.}\ }\textbf {\bibinfo {volume} {107}},\ \bibinfo {pages} {016401}
  (\bibinfo {year} {2011})}\BibitemShut {NoStop}%
\bibitem [{\citenamefont {Pardo}\ and\ \citenamefont
  {Pickett}(2009)}]{pardo2009half}%
  \BibitemOpen
  \bibfield  {author} {\bibinfo {author} {\bibfnamefont {V.}~\bibnamefont
  {Pardo}}\ and\ \bibinfo {author} {\bibfnamefont {W.~E.}\ \bibnamefont
  {Pickett}},\ }\href@noop {} {\bibfield  {journal} {\bibinfo  {journal} {Phys.
  Rev. Lett.}\ }\textbf {\bibinfo {volume} {102}},\ \bibinfo {pages} {166803}
  (\bibinfo {year} {2009})}\BibitemShut {NoStop}%
\bibitem [{\citenamefont {Qazilbash}\ \emph {et~al.}(2007)\citenamefont
  {Qazilbash}, \citenamefont {Brehm}, \citenamefont {Chae}, \citenamefont {Ho},
  \citenamefont {Andreev}, \citenamefont {Kim}, \citenamefont {Yun},
  \citenamefont {Balatsky}, \citenamefont {Maple}, \citenamefont {Keilmann}
  \emph {et~al.}}]{qazilbash2007mott}%
  \BibitemOpen
  \bibfield  {author} {\bibinfo {author} {\bibfnamefont {M.~M.}\ \bibnamefont
  {Qazilbash}}, \bibinfo {author} {\bibfnamefont {M.}~\bibnamefont {Brehm}},
  \bibinfo {author} {\bibfnamefont {B.-G.}\ \bibnamefont {Chae}}, \bibinfo
  {author} {\bibfnamefont {P.-C.}\ \bibnamefont {Ho}}, \bibinfo {author}
  {\bibfnamefont {G.~O.}\ \bibnamefont {Andreev}}, \bibinfo {author}
  {\bibfnamefont {B.-J.}\ \bibnamefont {Kim}}, \bibinfo {author} {\bibfnamefont
  {S.~J.}\ \bibnamefont {Yun}}, \bibinfo {author} {\bibfnamefont
  {A.}~\bibnamefont {Balatsky}}, \bibinfo {author} {\bibfnamefont
  {M.}~\bibnamefont {Maple}}, \bibinfo {author} {\bibfnamefont
  {F.}~\bibnamefont {Keilmann}},  \emph {et~al.},\ }\href@noop {} {\bibfield
  {journal} {\bibinfo  {journal} {Science}\ }\textbf {\bibinfo {volume}
  {318}},\ \bibinfo {pages} {1750} (\bibinfo {year} {2007})}\BibitemShut
  {NoStop}%
\bibitem [{\citenamefont {Haverkort}\ \emph {et~al.}(2005)\citenamefont
  {Haverkort}, \citenamefont {Hu}, \citenamefont {Tanaka}, \citenamefont
  {Reichelt}, \citenamefont {Streltsov}, \citenamefont {Korotin}, \citenamefont
  {Anisimov}, \citenamefont {Hsieh}, \citenamefont {Lin}, \citenamefont {Chen}
  \emph {et~al.}}]{haverkort2005orbital}%
  \BibitemOpen
  \bibfield  {author} {\bibinfo {author} {\bibfnamefont {M.}~\bibnamefont
  {Haverkort}}, \bibinfo {author} {\bibfnamefont {Z.}~\bibnamefont {Hu}},
  \bibinfo {author} {\bibfnamefont {A.}~\bibnamefont {Tanaka}}, \bibinfo
  {author} {\bibfnamefont {W.}~\bibnamefont {Reichelt}}, \bibinfo {author}
  {\bibfnamefont {S.}~\bibnamefont {Streltsov}}, \bibinfo {author}
  {\bibfnamefont {M.}~\bibnamefont {Korotin}}, \bibinfo {author} {\bibfnamefont
  {V.}~\bibnamefont {Anisimov}}, \bibinfo {author} {\bibfnamefont
  {H.}~\bibnamefont {Hsieh}}, \bibinfo {author} {\bibfnamefont {H.-J.}\
  \bibnamefont {Lin}}, \bibinfo {author} {\bibfnamefont {C.}~\bibnamefont
  {Chen}},  \emph {et~al.},\ }\href@noop {} {\bibfield  {journal} {\bibinfo
  {journal} {Phys. Rev. Lett.}\ }\textbf {\bibinfo {volume} {95}},\ \bibinfo
  {pages} {196404} (\bibinfo {year} {2005})}\BibitemShut {NoStop}%
\bibitem [{\citenamefont {Berglund}\ and\ \citenamefont
  {Guggenheim}(1969)}]{berglund1969electronic}%
  \BibitemOpen
  \bibfield  {author} {\bibinfo {author} {\bibfnamefont {C.}~\bibnamefont
  {Berglund}}\ and\ \bibinfo {author} {\bibfnamefont {H.}~\bibnamefont
  {Guggenheim}},\ }\href@noop {} {\bibfield  {journal} {\bibinfo  {journal}
  {Phys. Rev.}\ }\textbf {\bibinfo {volume} {185}},\ \bibinfo {pages} {1022}
  (\bibinfo {year} {1969})}\BibitemShut {NoStop}%
\bibitem [{\citenamefont {Berglund}\ and\ \citenamefont
  {Jayaraman}(1969)}]{berglund1969hydrostatic}%
  \BibitemOpen
  \bibfield  {author} {\bibinfo {author} {\bibfnamefont {C.}~\bibnamefont
  {Berglund}}\ and\ \bibinfo {author} {\bibfnamefont {A.}~\bibnamefont
  {Jayaraman}},\ }\href@noop {} {\bibfield  {journal} {\bibinfo  {journal}
  {Phys. Rev.}\ }\textbf {\bibinfo {volume} {185}},\ \bibinfo {pages} {1034}
  (\bibinfo {year} {1969})}\BibitemShut {NoStop}%
\bibitem [{\citenamefont {Goodenough}(1971)}]{goodenough1971two}%
  \BibitemOpen
  \bibfield  {author} {\bibinfo {author} {\bibfnamefont {J.~B.}\ \bibnamefont
  {Goodenough}},\ }\href@noop {} {\bibfield  {journal} {\bibinfo  {journal} {J.
  Solid State Chem.}\ }\textbf {\bibinfo {volume} {3}},\ \bibinfo {pages} {490}
  (\bibinfo {year} {1971})}\BibitemShut {NoStop}%
\bibitem [{\citenamefont {Morin}(1959)}]{morin1959oxides}%
  \BibitemOpen
  \bibfield  {author} {\bibinfo {author} {\bibfnamefont {F.}~\bibnamefont
  {Morin}},\ }\href@noop {} {\bibfield  {journal} {\bibinfo  {journal} {Phys.
  Rev. Lett.}\ }\textbf {\bibinfo {volume} {3}},\ \bibinfo {pages} {34}
  (\bibinfo {year} {1959})}\BibitemShut {NoStop}%
\bibitem [{\citenamefont {Laverock}\ \emph {et~al.}(2012)\citenamefont
  {Laverock}, \citenamefont {Piper}, \citenamefont {Preston}, \citenamefont
  {Chen}, \citenamefont {McNulty}, \citenamefont {Smith}, \citenamefont
  {Kittiwatanakul}, \citenamefont {Lu}, \citenamefont {Wolf}, \citenamefont
  {Glans} \emph {et~al.}}]{laverock2012strain}%
  \BibitemOpen
  \bibfield  {author} {\bibinfo {author} {\bibfnamefont {J.}~\bibnamefont
  {Laverock}}, \bibinfo {author} {\bibfnamefont {L.}~\bibnamefont {Piper}},
  \bibinfo {author} {\bibfnamefont {A.}~\bibnamefont {Preston}}, \bibinfo
  {author} {\bibfnamefont {B.}~\bibnamefont {Chen}}, \bibinfo {author}
  {\bibfnamefont {J.}~\bibnamefont {McNulty}}, \bibinfo {author} {\bibfnamefont
  {K.}~\bibnamefont {Smith}}, \bibinfo {author} {\bibfnamefont
  {S.}~\bibnamefont {Kittiwatanakul}}, \bibinfo {author} {\bibfnamefont
  {J.}~\bibnamefont {Lu}}, \bibinfo {author} {\bibfnamefont {S.}~\bibnamefont
  {Wolf}}, \bibinfo {author} {\bibfnamefont {P.-A.}\ \bibnamefont {Glans}},
  \emph {et~al.},\ }\href@noop {} {\bibfield  {journal} {\bibinfo  {journal}
  {Phys. Rev. B}\ }\textbf {\bibinfo {volume} {85}},\ \bibinfo {pages} {081104}
  (\bibinfo {year} {2012})}\BibitemShut {NoStop}%
\bibitem [{\citenamefont {Lazarovits}\ \emph {et~al.}(2010)\citenamefont
  {Lazarovits}, \citenamefont {Kim}, \citenamefont {Haule},\ and\ \citenamefont
  {Kotliar}}]{lazarovits2010effects}%
  \BibitemOpen
  \bibfield  {author} {\bibinfo {author} {\bibfnamefont {B.}~\bibnamefont
  {Lazarovits}}, \bibinfo {author} {\bibfnamefont {K.}~\bibnamefont {Kim}},
  \bibinfo {author} {\bibfnamefont {K.}~\bibnamefont {Haule}}, \ and\ \bibinfo
  {author} {\bibfnamefont {G.}~\bibnamefont {Kotliar}},\ }\href@noop {}
  {\bibfield  {journal} {\bibinfo  {journal} {Phys. Rev. B}\ }\textbf {\bibinfo
  {volume} {81}},\ \bibinfo {pages} {115117} (\bibinfo {year}
  {2010})}\BibitemShut {NoStop}%
\bibitem [{\citenamefont {Muraoka}, \citenamefont {Ueda},\ and\ \citenamefont
  {Hiroi}(2002)}]{muraoka2002large}%
  \BibitemOpen
  \bibfield  {author} {\bibinfo {author} {\bibfnamefont {Y.}~\bibnamefont
  {Muraoka}}, \bibinfo {author} {\bibfnamefont {Y.}~\bibnamefont {Ueda}}, \
  and\ \bibinfo {author} {\bibfnamefont {Z.}~\bibnamefont {Hiroi}},\
  }\href@noop {} {\bibfield  {journal} {\bibinfo  {journal} {J. Phys. Chem.
  Solids}\ }\textbf {\bibinfo {volume} {63}},\ \bibinfo {pages} {965} (\bibinfo
  {year} {2002})}\BibitemShut {NoStop}%
\bibitem [{\citenamefont {Nakano}\ \emph {et~al.}(2012)\citenamefont {Nakano},
  \citenamefont {Shibuya}, \citenamefont {Okuyama}, \citenamefont {Hatano},
  \citenamefont {Ono}, \citenamefont {Kawasaki}, \citenamefont {Iwasa},\ and\
  \citenamefont {Tokura}}]{nakano2012collective}%
  \BibitemOpen
  \bibfield  {author} {\bibinfo {author} {\bibfnamefont {M.}~\bibnamefont
  {Nakano}}, \bibinfo {author} {\bibfnamefont {K.}~\bibnamefont {Shibuya}},
  \bibinfo {author} {\bibfnamefont {D.}~\bibnamefont {Okuyama}}, \bibinfo
  {author} {\bibfnamefont {T.}~\bibnamefont {Hatano}}, \bibinfo {author}
  {\bibfnamefont {S.}~\bibnamefont {Ono}}, \bibinfo {author} {\bibfnamefont
  {M.}~\bibnamefont {Kawasaki}}, \bibinfo {author} {\bibfnamefont
  {Y.}~\bibnamefont {Iwasa}}, \ and\ \bibinfo {author} {\bibfnamefont
  {Y.}~\bibnamefont {Tokura}},\ }\href@noop {} {\bibfield  {journal} {\bibinfo
  {journal} {Nature}\ }\textbf {\bibinfo {volume} {487}},\ \bibinfo {pages}
  {459} (\bibinfo {year} {2012})}\BibitemShut {NoStop}%
\bibitem [{\citenamefont {Okazaki}\ \emph {et~al.}(2004)\citenamefont
  {Okazaki}, \citenamefont {Wadati}, \citenamefont {Fujimori}, \citenamefont
  {Onoda}, \citenamefont {Muraoka},\ and\ \citenamefont
  {Hiroi}}]{okazaki2004photoemission}%
  \BibitemOpen
  \bibfield  {author} {\bibinfo {author} {\bibfnamefont {K.}~\bibnamefont
  {Okazaki}}, \bibinfo {author} {\bibfnamefont {H.}~\bibnamefont {Wadati}},
  \bibinfo {author} {\bibfnamefont {A.}~\bibnamefont {Fujimori}}, \bibinfo
  {author} {\bibfnamefont {M.}~\bibnamefont {Onoda}}, \bibinfo {author}
  {\bibfnamefont {Y.}~\bibnamefont {Muraoka}}, \ and\ \bibinfo {author}
  {\bibfnamefont {Z.}~\bibnamefont {Hiroi}},\ }\href@noop {} {\bibfield
  {journal} {\bibinfo  {journal} {Phys. Rev. B}\ }\textbf {\bibinfo {volume}
  {69}},\ \bibinfo {pages} {165104} (\bibinfo {year} {2004})}\BibitemShut
  {NoStop}%
\bibitem [{\citenamefont {Newns}\ \emph {et~al.}(1998)\citenamefont {Newns},
  \citenamefont {Misewich}, \citenamefont {Tsuei}, \citenamefont {Gupta},
  \citenamefont {Scott},\ and\ \citenamefont {Schrott}}]{newns1998mott}%
  \BibitemOpen
  \bibfield  {author} {\bibinfo {author} {\bibfnamefont {D.}~\bibnamefont
  {Newns}}, \bibinfo {author} {\bibfnamefont {J.}~\bibnamefont {Misewich}},
  \bibinfo {author} {\bibfnamefont {C.}~\bibnamefont {Tsuei}}, \bibinfo
  {author} {\bibfnamefont {A.}~\bibnamefont {Gupta}}, \bibinfo {author}
  {\bibfnamefont {B.}~\bibnamefont {Scott}}, \ and\ \bibinfo {author}
  {\bibfnamefont {A.}~\bibnamefont {Schrott}},\ }\href@noop {} {\bibfield
  {journal} {\bibinfo  {journal} {Appl. Phys. Lett.}\ }\textbf {\bibinfo
  {volume} {73}},\ \bibinfo {pages} {780} (\bibinfo {year} {1998})}\BibitemShut
  {NoStop}%
\bibitem [{\citenamefont {Yajima}, \citenamefont {Nishimura},\ and\
  \citenamefont {Toriumi}(2015)}]{yajima2015positive}%
  \BibitemOpen
  \bibfield  {author} {\bibinfo {author} {\bibfnamefont {T.}~\bibnamefont
  {Yajima}}, \bibinfo {author} {\bibfnamefont {T.}~\bibnamefont {Nishimura}}, \
  and\ \bibinfo {author} {\bibfnamefont {A.}~\bibnamefont {Toriumi}},\
  }\href@noop {} {\bibfield  {journal} {\bibinfo  {journal} {Nat. Commun.}\
  }\textbf {\bibinfo {volume} {6}},\ \bibinfo {pages} {10104} (\bibinfo {year}
  {2015})}\BibitemShut {NoStop}%
\bibitem [{\citenamefont {Muraoka}\ and\ \citenamefont
  {Hiroi}(2002)}]{muraoka2002metal}%
  \BibitemOpen
  \bibfield  {author} {\bibinfo {author} {\bibfnamefont {Y.}~\bibnamefont
  {Muraoka}}\ and\ \bibinfo {author} {\bibfnamefont {Z.}~\bibnamefont
  {Hiroi}},\ }\href@noop {} {\bibfield  {journal} {\bibinfo  {journal} {Appl.
  Phys. Lett.}\ }\textbf {\bibinfo {volume} {80}},\ \bibinfo {pages} {583}
  (\bibinfo {year} {2002})}\BibitemShut {NoStop}%
\bibitem [{\citenamefont {Aetukuri}\ \emph {et~al.}(2013)\citenamefont
  {Aetukuri}, \citenamefont {Gray}, \citenamefont {Drouard}, \citenamefont
  {Cossale}, \citenamefont {Gao}, \citenamefont {Reid}, \citenamefont
  {Kukreja}, \citenamefont {Ohldag}, \citenamefont {Jenkins}, \citenamefont
  {Arenholz} \emph {et~al.}}]{aetukuri2013control}%
  \BibitemOpen
  \bibfield  {author} {\bibinfo {author} {\bibfnamefont {N.~B.}\ \bibnamefont
  {Aetukuri}}, \bibinfo {author} {\bibfnamefont {A.~X.}\ \bibnamefont {Gray}},
  \bibinfo {author} {\bibfnamefont {M.}~\bibnamefont {Drouard}}, \bibinfo
  {author} {\bibfnamefont {M.}~\bibnamefont {Cossale}}, \bibinfo {author}
  {\bibfnamefont {L.}~\bibnamefont {Gao}}, \bibinfo {author} {\bibfnamefont
  {A.~H.}\ \bibnamefont {Reid}}, \bibinfo {author} {\bibfnamefont
  {R.}~\bibnamefont {Kukreja}}, \bibinfo {author} {\bibfnamefont
  {H.}~\bibnamefont {Ohldag}}, \bibinfo {author} {\bibfnamefont {C.~A.}\
  \bibnamefont {Jenkins}}, \bibinfo {author} {\bibfnamefont {E.}~\bibnamefont
  {Arenholz}},  \emph {et~al.},\ }\href@noop {} {\bibfield  {journal} {\bibinfo
   {journal} {Nat. Phys.}\ }\textbf {\bibinfo {volume} {9}},\ \bibinfo {pages}
  {661} (\bibinfo {year} {2013})}\BibitemShut {NoStop}%
\bibitem [{\citenamefont {Jeong}\ \emph {et~al.}(2013)\citenamefont {Jeong},
  \citenamefont {Aetukuri}, \citenamefont {Graf}, \citenamefont {Schladt},
  \citenamefont {Samant},\ and\ \citenamefont {Parkin}}]{jeong2013suppression}%
  \BibitemOpen
  \bibfield  {author} {\bibinfo {author} {\bibfnamefont {J.}~\bibnamefont
  {Jeong}}, \bibinfo {author} {\bibfnamefont {N.}~\bibnamefont {Aetukuri}},
  \bibinfo {author} {\bibfnamefont {T.}~\bibnamefont {Graf}}, \bibinfo {author}
  {\bibfnamefont {T.~D.}\ \bibnamefont {Schladt}}, \bibinfo {author}
  {\bibfnamefont {M.~G.}\ \bibnamefont {Samant}}, \ and\ \bibinfo {author}
  {\bibfnamefont {S.~S.}\ \bibnamefont {Parkin}},\ }\href@noop {} {\bibfield
  {journal} {\bibinfo  {journal} {Science}\ }\textbf {\bibinfo {volume}
  {339}},\ \bibinfo {pages} {1402} (\bibinfo {year} {2013})}\BibitemShut
  {NoStop}%
\bibitem [{\citenamefont {Ramamoorthy}, \citenamefont {Vanderbilt},\ and\
  \citenamefont {King-Smith}(1994)}]{ramamoorthy1994first}%
  \BibitemOpen
  \bibfield  {author} {\bibinfo {author} {\bibfnamefont {M.}~\bibnamefont
  {Ramamoorthy}}, \bibinfo {author} {\bibfnamefont {D.}~\bibnamefont
  {Vanderbilt}}, \ and\ \bibinfo {author} {\bibfnamefont {R.}~\bibnamefont
  {King-Smith}},\ }\href@noop {} {\bibfield  {journal} {\bibinfo  {journal}
  {Phys. Rev. B}\ }\textbf {\bibinfo {volume} {49}},\ \bibinfo {pages} {16721}
  (\bibinfo {year} {1994})}\BibitemShut {NoStop}%
\bibitem [{\citenamefont {Diebold}(2003)}]{diebold2003surface}%
  \BibitemOpen
  \bibfield  {author} {\bibinfo {author} {\bibfnamefont {U.}~\bibnamefont
  {Diebold}},\ }\href@noop {} {\bibfield  {journal} {\bibinfo  {journal} {Surf.
  Sci. Rep.}\ }\textbf {\bibinfo {volume} {48}},\ \bibinfo {pages} {53}
  (\bibinfo {year} {2003})}\BibitemShut {NoStop}%
\bibitem [{\citenamefont {Aetukuri}()}]{phanithesis}%
  \BibitemOpen
  \bibfield  {author} {\bibinfo {author} {\bibfnamefont {N.~P.}\ \bibnamefont
  {Aetukuri}},\ }\href@noop {} {\bibinfo  {journal} {Ph.D. thesis, Stanford
  University, 2013}\ }\BibitemShut {NoStop}%
\bibitem [{\citenamefont {Li}\ and\ \citenamefont
  {Dho}(2014)}]{li2014characteristics}%
  \BibitemOpen
\bibfield  {journal} {  }\bibfield  {author} {\bibinfo {author} {\bibfnamefont
  {J.}~\bibnamefont {Li}}\ and\ \bibinfo {author} {\bibfnamefont
  {J.}~\bibnamefont {Dho}},\ }\href@noop {} {\bibfield  {journal} {\bibinfo
  {journal} {J. Cryst. Growth}\ }\textbf {\bibinfo {volume} {404}},\ \bibinfo
  {pages} {84} (\bibinfo {year} {2014})}\BibitemShut {NoStop}%
\bibitem [{\citenamefont {Jeong}\ \emph {et~al.}(2015)\citenamefont {Jeong},
  \citenamefont {Aetukuri}, \citenamefont {Passarello}, \citenamefont
  {Conradson}, \citenamefont {Samant},\ and\ \citenamefont
  {Parkin}}]{jeong2015giant}%
  \BibitemOpen
  \bibfield  {author} {\bibinfo {author} {\bibfnamefont {J.}~\bibnamefont
  {Jeong}}, \bibinfo {author} {\bibfnamefont {N.~B.}\ \bibnamefont {Aetukuri}},
  \bibinfo {author} {\bibfnamefont {D.}~\bibnamefont {Passarello}}, \bibinfo
  {author} {\bibfnamefont {S.~D.}\ \bibnamefont {Conradson}}, \bibinfo {author}
  {\bibfnamefont {M.~G.}\ \bibnamefont {Samant}}, \ and\ \bibinfo {author}
  {\bibfnamefont {S.~S.}\ \bibnamefont {Parkin}},\ }\href@noop {} {\bibfield
  {journal} {\bibinfo  {journal} {Proc. Natl. Acad. Sci. U.S.A.}\ }\textbf
  {\bibinfo {volume} {112}},\ \bibinfo {pages} {1013} (\bibinfo {year}
  {2015})}\BibitemShut {NoStop}%
\bibitem [{\citenamefont {Martens}\ \emph {et~al.}(2014)\citenamefont
  {Martens}, \citenamefont {Aetukuri}, \citenamefont {Jeong}, \citenamefont
  {Samant},\ and\ \citenamefont {Parkin}}]{martens2014improved}%
  \BibitemOpen
  \bibfield  {author} {\bibinfo {author} {\bibfnamefont {K.}~\bibnamefont
  {Martens}}, \bibinfo {author} {\bibfnamefont {N.}~\bibnamefont {Aetukuri}},
  \bibinfo {author} {\bibfnamefont {J.}~\bibnamefont {Jeong}}, \bibinfo
  {author} {\bibfnamefont {M.~G.}\ \bibnamefont {Samant}}, \ and\ \bibinfo
  {author} {\bibfnamefont {S.~S.}\ \bibnamefont {Parkin}},\ }\href@noop {}
  {\bibfield  {journal} {\bibinfo  {journal} {Appl. Phys. Lett.}\ }\textbf
  {\bibinfo {volume} {104}},\ \bibinfo {pages} {081918} (\bibinfo {year}
  {2014})}\BibitemShut {NoStop}%
\bibitem [{\citenamefont {Gilmer}\ and\ \citenamefont
  {Grabow}(1987)}]{gilmer1987models}%
  \BibitemOpen
  \bibfield  {author} {\bibinfo {author} {\bibfnamefont {G.~H.}\ \bibnamefont
  {Gilmer}}\ and\ \bibinfo {author} {\bibfnamefont {M.~H.}\ \bibnamefont
  {Grabow}},\ }\href@noop {} {\bibfield  {journal} {\bibinfo  {journal} {JOM}\
  }\textbf {\bibinfo {volume} {39}},\ \bibinfo {pages} {19} (\bibinfo {year}
  {1987})}\BibitemShut {NoStop}%
\bibitem [{\citenamefont {Hong}\ \emph {et~al.}(2005)\citenamefont {Hong},
  \citenamefont {Lee}, \citenamefont {Yoon}, \citenamefont {Christen},
  \citenamefont {Lowndes}, \citenamefont {Suo},\ and\ \citenamefont
  {Zhang}}]{hong2005persistent}%
  \BibitemOpen
  \bibfield  {author} {\bibinfo {author} {\bibfnamefont {W.}~\bibnamefont
  {Hong}}, \bibinfo {author} {\bibfnamefont {H.~N.}\ \bibnamefont {Lee}},
  \bibinfo {author} {\bibfnamefont {M.}~\bibnamefont {Yoon}}, \bibinfo {author}
  {\bibfnamefont {H.~M.}\ \bibnamefont {Christen}}, \bibinfo {author}
  {\bibfnamefont {D.~H.}\ \bibnamefont {Lowndes}}, \bibinfo {author}
  {\bibfnamefont {Z.}~\bibnamefont {Suo}}, \ and\ \bibinfo {author}
  {\bibfnamefont {Z.}~\bibnamefont {Zhang}},\ }\href@noop {} {\bibfield
  {journal} {\bibinfo  {journal} {Phys. Rev. Lett.}\ }\textbf {\bibinfo
  {volume} {95}},\ \bibinfo {pages} {095501} (\bibinfo {year}
  {2005})}\BibitemShut {NoStop}%
\bibitem [{\citenamefont {Koster}\ \emph {et~al.}(1999)\citenamefont {Koster},
  \citenamefont {Rijnders}, \citenamefont {Blank},\ and\ \citenamefont
  {Rogalla}}]{koster1999imposed}%
  \BibitemOpen
  \bibfield  {author} {\bibinfo {author} {\bibfnamefont {G.}~\bibnamefont
  {Koster}}, \bibinfo {author} {\bibfnamefont {G.~J.}\ \bibnamefont
  {Rijnders}}, \bibinfo {author} {\bibfnamefont {D.~H.}\ \bibnamefont {Blank}},
  \ and\ \bibinfo {author} {\bibfnamefont {H.}~\bibnamefont {Rogalla}},\
  }\href@noop {} {\bibfield  {journal} {\bibinfo  {journal} {Appl. phys.
  lett.}\ }\textbf {\bibinfo {volume} {74}},\ \bibinfo {pages} {3729} (\bibinfo
  {year} {1999})}\BibitemShut {NoStop}%
\bibitem [{\citenamefont {Liu}\ \emph {et~al.}(2012)\citenamefont {Liu},
  \citenamefont {Hwang}, \citenamefont {Tao}, \citenamefont {Strikwerda},
  \citenamefont {Fan}, \citenamefont {Keiser}, \citenamefont {Sternbach},
  \citenamefont {West}, \citenamefont {Kittiwatanakul}, \citenamefont {Lu}
  \emph {et~al.}}]{liu2012terahertz}%
  \BibitemOpen
  \bibfield  {author} {\bibinfo {author} {\bibfnamefont {M.}~\bibnamefont
  {Liu}}, \bibinfo {author} {\bibfnamefont {H.~Y.}\ \bibnamefont {Hwang}},
  \bibinfo {author} {\bibfnamefont {H.}~\bibnamefont {Tao}}, \bibinfo {author}
  {\bibfnamefont {A.~C.}\ \bibnamefont {Strikwerda}}, \bibinfo {author}
  {\bibfnamefont {K.}~\bibnamefont {Fan}}, \bibinfo {author} {\bibfnamefont
  {G.~R.}\ \bibnamefont {Keiser}}, \bibinfo {author} {\bibfnamefont {A.~J.}\
  \bibnamefont {Sternbach}}, \bibinfo {author} {\bibfnamefont {K.~G.}\
  \bibnamefont {West}}, \bibinfo {author} {\bibfnamefont {S.}~\bibnamefont
  {Kittiwatanakul}}, \bibinfo {author} {\bibfnamefont {J.}~\bibnamefont {Lu}},
  \emph {et~al.},\ }\href@noop {} {\bibfield  {journal} {\bibinfo  {journal}
  {Nature}\ }\textbf {\bibinfo {volume} {487}},\ \bibinfo {pages} {345}
  (\bibinfo {year} {2012})}\BibitemShut {NoStop}%
\bibitem [{\citenamefont {Dicken}\ \emph {et~al.}(2009)\citenamefont {Dicken},
  \citenamefont {Aydin}, \citenamefont {Pryce}, \citenamefont {Sweatlock},
  \citenamefont {Boyd}, \citenamefont {Walavalkar}, \citenamefont {Ma},\ and\
  \citenamefont {Atwater}}]{dicken2009frequency}%
  \BibitemOpen
  \bibfield  {author} {\bibinfo {author} {\bibfnamefont {M.~J.}\ \bibnamefont
  {Dicken}}, \bibinfo {author} {\bibfnamefont {K.}~\bibnamefont {Aydin}},
  \bibinfo {author} {\bibfnamefont {I.~M.}\ \bibnamefont {Pryce}}, \bibinfo
  {author} {\bibfnamefont {L.~A.}\ \bibnamefont {Sweatlock}}, \bibinfo {author}
  {\bibfnamefont {E.~M.}\ \bibnamefont {Boyd}}, \bibinfo {author}
  {\bibfnamefont {S.}~\bibnamefont {Walavalkar}}, \bibinfo {author}
  {\bibfnamefont {J.}~\bibnamefont {Ma}}, \ and\ \bibinfo {author}
  {\bibfnamefont {H.~A.}\ \bibnamefont {Atwater}},\ }\href@noop {} {\bibfield
  {journal} {\bibinfo  {journal} {Opt. Express}\ }\textbf {\bibinfo {volume}
  {17}},\ \bibinfo {pages} {18330} (\bibinfo {year} {2009})}\BibitemShut
  {NoStop}%
\bibitem [{\citenamefont {Kabashin}\ \emph {et~al.}(2009)\citenamefont
  {Kabashin}, \citenamefont {Evans}, \citenamefont {Pastkovsky}, \citenamefont
  {Hendren}, \citenamefont {Wurtz}, \citenamefont {Atkinson}, \citenamefont
  {Pollard}, \citenamefont {Podolskiy},\ and\ \citenamefont
  {Zayats}}]{kabashin2009plasmonic}%
  \BibitemOpen
  \bibfield  {author} {\bibinfo {author} {\bibfnamefont {A.}~\bibnamefont
  {Kabashin}}, \bibinfo {author} {\bibfnamefont {P.}~\bibnamefont {Evans}},
  \bibinfo {author} {\bibfnamefont {S.}~\bibnamefont {Pastkovsky}}, \bibinfo
  {author} {\bibfnamefont {W.}~\bibnamefont {Hendren}}, \bibinfo {author}
  {\bibfnamefont {G.}~\bibnamefont {Wurtz}}, \bibinfo {author} {\bibfnamefont
  {R.}~\bibnamefont {Atkinson}}, \bibinfo {author} {\bibfnamefont
  {R.}~\bibnamefont {Pollard}}, \bibinfo {author} {\bibfnamefont
  {V.}~\bibnamefont {Podolskiy}}, \ and\ \bibinfo {author} {\bibfnamefont
  {A.}~\bibnamefont {Zayats}},\ }\href@noop {} {\bibfield  {journal} {\bibinfo
  {journal} {Nat. Mater.}\ }\textbf {\bibinfo {volume} {8}},\ \bibinfo {pages}
  {867} (\bibinfo {year} {2009})}\BibitemShut {NoStop}%
\bibitem [{\citenamefont {{Ichimiya}}\ and\ \citenamefont
  {{Cohen}}(2004)}]{Ichimiya2004RHEED}%
  \BibitemOpen
  \bibfield  {author} {\bibinfo {author} {\bibfnamefont {A.}~\bibnamefont
  {{Ichimiya}}}\ and\ \bibinfo {author} {\bibfnamefont {P.~I.}\ \bibnamefont
  {{Cohen}}},\ }\href@noop {} {\emph {\bibinfo {title} {{Reflection High-Energy
  Electron Diffraction}}}}\ (\bibinfo  {publisher} {Cambridge University Press,
  Cambridge},\ \bibinfo {year} {2004})\BibitemShut {NoStop}%
\bibitem [{\citenamefont {Kawatani}, \citenamefont {Kanki},\ and\ \citenamefont
  {Tanaka}(2014)}]{kawatani2014formation}%
  \BibitemOpen
  \bibfield  {author} {\bibinfo {author} {\bibfnamefont {K.}~\bibnamefont
  {Kawatani}}, \bibinfo {author} {\bibfnamefont {T.}~\bibnamefont {Kanki}}, \
  and\ \bibinfo {author} {\bibfnamefont {H.}~\bibnamefont {Tanaka}},\
  }\href@noop {} {\bibfield  {journal} {\bibinfo  {journal} {Phys. Rev. B}\
  }\textbf {\bibinfo {volume} {90}},\ \bibinfo {pages} {054203} (\bibinfo
  {year} {2014})}\BibitemShut {NoStop}%
\bibitem [{\citenamefont {Cai}\ \emph {et~al.}(2015)\citenamefont {Cai},
  \citenamefont {Li}, \citenamefont {Wang}, \citenamefont {Ju}, \citenamefont
  {Feng},\ and\ \citenamefont {Gong}}]{cai2015single}%
  \BibitemOpen
  \bibfield  {author} {\bibinfo {author} {\bibfnamefont {T.}~\bibnamefont
  {Cai}}, \bibinfo {author} {\bibfnamefont {X.}~\bibnamefont {Li}}, \bibinfo
  {author} {\bibfnamefont {F.}~\bibnamefont {Wang}}, \bibinfo {author}
  {\bibfnamefont {S.}~\bibnamefont {Ju}}, \bibinfo {author} {\bibfnamefont
  {J.}~\bibnamefont {Feng}}, \ and\ \bibinfo {author} {\bibfnamefont {C.-D.}\
  \bibnamefont {Gong}},\ }\href@noop {} {\bibfield  {journal} {\bibinfo
  {journal} {Nano. lett.}\ }\textbf {\bibinfo {volume} {15}},\ \bibinfo {pages}
  {6434} (\bibinfo {year} {2015})}\BibitemShut {NoStop}%
\end{thebibliography}%


\begin{thebibliography}{1}%
\makeatletter
\providecommand \@ifxundefined [1]{%
 \@ifx{#1\undefined}
}%
\providecommand \@ifnum [1]{%
 \ifnum #1\expandafter \@firstoftwo
 \else \expandafter \@secondoftwo
 \fi
}%
\providecommand \@ifx [1]{%
 \ifx #1\expandafter \@firstoftwo
 \else \expandafter \@secondoftwo
 \fi
}%
\providecommand \natexlab [1]{#1}%
\providecommand \enquote  [1]{``#1''}%
\providecommand \bibnamefont  [1]{#1}%
\providecommand \bibfnamefont [1]{#1}%
\providecommand \citenamefont [1]{#1}%
\providecommand \href@noop [0]{\@secondoftwo}%
\providecommand \href [0]{\begingroup \@sanitize@url \@href}%
\providecommand \@href[1]{\@@startlink{#1}\@@href}%
\providecommand \@@href[1]{\endgroup#1\@@endlink}%
\providecommand \@sanitize@url [0]{\catcode `\\12\catcode `\$12\catcode
  `\&12\catcode `\#12\catcode `\^12\catcode `\_12\catcode `\%12\relax}%
\providecommand \@@startlink[1]{}%
\providecommand \@@endlink[0]{}%
\providecommand \url  [0]{\begingroup\@sanitize@url \@url }%
\providecommand \@url [1]{\endgroup\@href {#1}{\urlprefix }}%
\providecommand \urlprefix  [0]{URL }%
\providecommand \Eprint [0]{\href }%
\providecommand \doibase [0]{http://dx.doi.org/}%
\providecommand \selectlanguage [0]{\@gobble}%
\providecommand \bibinfo  [0]{\@secondoftwo}%
\providecommand \bibfield  [0]{\@secondoftwo}%
\providecommand \translation [1]{[#1]}%
\providecommand \BibitemOpen [0]{}%
\providecommand \bibitemStop [0]{}%
\providecommand \bibitemNoStop [0]{.\EOS\space}%
\providecommand \EOS [0]{\spacefactor3000\relax}%
\providecommand \BibitemShut  [1]{\csname bibitem#1\endcsname}%
\let\auto@bib@innerbib\@empty
\bibitem [{\citenamefont {Aetukuri}\ \emph {et~al.}(2013)\citenamefont
  {Aetukuri}, \citenamefont {Gray}, \citenamefont {Drouard}, \citenamefont
  {Cossale}, \citenamefont {Gao}, \citenamefont {Reid}, \citenamefont
  {Kukreja}, \citenamefont {Ohldag}, \citenamefont {Jenkins}, \citenamefont
  {Arenholz} \emph {et~al.}}]{aetukuri2013control}%
  \BibitemOpen
  \bibfield  {author} {\bibinfo {author} {\bibfnamefont {N.~B.}\ \bibnamefont
  {Aetukuri}}, \bibinfo {author} {\bibfnamefont {A.~X.}\ \bibnamefont {Gray}},
  \bibinfo {author} {\bibfnamefont {M.}~\bibnamefont {Drouard}}, \bibinfo
  {author} {\bibfnamefont {M.}~\bibnamefont {Cossale}}, \bibinfo {author}
  {\bibfnamefont {L.}~\bibnamefont {Gao}}, \bibinfo {author} {\bibfnamefont
  {A.~H.}\ \bibnamefont {Reid}}, \bibinfo {author} {\bibfnamefont
  {R.}~\bibnamefont {Kukreja}}, \bibinfo {author} {\bibfnamefont
  {H.}~\bibnamefont {Ohldag}}, \bibinfo {author} {\bibfnamefont {C.~A.}\
  \bibnamefont {Jenkins}}, \bibinfo {author} {\bibfnamefont {E.}~\bibnamefont
  {Arenholz}},  \emph {et~al.},\ }\href@noop {} {\bibfield  {journal} {\bibinfo
   {journal} {Nat. Phys.}\ }\textbf {\bibinfo {volume} {9}},\ \bibinfo {pages}
  {661} (\bibinfo {year} {2013})}\BibitemShut {NoStop}%
\end{thebibliography}%
\end{document}


\preprint{AIP/123-QED}

\title{Supplementary Material for "Atomically-Smooth Single-Crystalline VO$_2$ thin films with Bulk-like Metal-Insulator Transitions"}

\author{Debasish Mondal}
\altaffiliation{These authors contributed equally}
\affiliation{Solid State and Structural Chemistry Unit, Indian institute of Science, Bengaluru, India}

\author{Smruti Rekha Mahapatra}
\altaffiliation{These authors contributed equally}
\affiliation{Solid State and Structural Chemistry Unit, Indian institute of Science, Bengaluru, India}

\author{Tanweer Ahmed}
\affiliation{Department of Physics, Indian institute of Science, Bengaluru, India}

\author{Suresh Kumar Podapangi}
\affiliation{Solid State and Structural Chemistry Unit, Indian institute of Science, Bengaluru, India}

\author{Arindam Ghosh}
\affiliation{Department of Physics, Indian institute of Science, Bengaluru, India}

\author{Naga Phani B. Aetukuri}
\email{phani@alumni.stanford.edu; phani@iisc.ac.in}
\affiliation{Solid State and Structural Chemistry Unit, Indian institute of Science, Bengaluru, India}

\date{\today}
	\renewcommand{\thefigure}{S\arabic{figure}}%
\maketitle
\onecolumngrid
\section{O\MakeLowercase{ut of plane strain calculation for} VO$_2$ \MakeLowercase{${(21\bar{1})_M}$ plane on} T\MakeLowercase{i}O$_2$ \MakeLowercase{${(101)_R}$ plane:}}
The stress-strain relationship along the three mutually perpendicular axes is given by the following equations.${^[}$\cite{aetukuri2013control}${^]}$
\begin{eqnarray}
\epsilon_x =\frac{1}{E}{[{\sigma_x}-v({\sigma_y}+{\sigma_z})]},\\
\epsilon_y =\frac{1}{E}{[\sigma_y-v(\sigma_x+\sigma_z)]},\\
\epsilon_z =\frac{1}{E}{[\sigma_z-v(\sigma_x+\sigma_y)]},
\end{eqnarray}
where ${\sigma_x}$, ${\sigma_y}$, ${\sigma_z}$ are the stresses along x, y, z axes respectively and ${\epsilon_x}$, ${\epsilon_y}$, ${\epsilon_z}$ are the resultant strains due to the stresses along x, y, z axes respectively. E and v are the Young’s modulus and Poisson’s ratio of the material respectively. As there is no stress along z axis, i.e. ${\sigma_z} = 0$, modified stress-strain relationships can be rewritten as follows.${^[}$\cite{aetukuri2013control}${^]}$
\begin{eqnarray}
\epsilon_x=\frac{1}{E}{[\sigma_x-v\sigma_y]},\\
\epsilon_y=\frac{1}{E}{[\sigma_y-v\sigma_x]},\\
\epsilon_z=\frac{1}{E}{[-v(\sigma_x + \sigma_y)]},
\end{eqnarray}
From, eq. (4) and eq. (5), it can be written
\begin{eqnarray}
\epsilon_x + \epsilon_y = \frac{(\sigma_x + \sigma_y)}{E}(1 - v)
\end{eqnarray}
From eq. (6) and eq. (7), out of plane strain ($\epsilon_z$) can be written as follows
\begin{eqnarray}
\epsilon_z = -\frac{v(\epsilon_x + \epsilon_y)}{(1 - v)}
\end{eqnarray}
Now strain can be defined as,
\begin{eqnarray}
\epsilon_x = \frac{x_M^{film} - x_M^{bulk}}{x_M^{bulk}},\\
\epsilon_y = \frac{y_M^{film} - y_M^{bulk}}{y_M^{bulk}},\\
\epsilon_z = \frac{z_M^{film} - z_M^{bulk}}{z_M^{bulk}},
\end{eqnarray}
For, fully-strained film, ${x_M^{film}}$ = ${\sqrt{a_{R(TiO_2)}^2 + c_{R(TiO_2)}^2}}$ = 5.4645 \AA, ${x_M^{bulk}}$ =  ${\sqrt{\frac{1}{4}a_{M(VO_2)}^2 + b_{M(VO_2)}^2}}$ = 5.3723 \AA,  ${y_M^{film}}$ = ${b_{R(TiO_2)}}$ = 4.5941 \AA, ${y_M^{bulk}}$ = ${\sqrt{c_{M(VO_2)}^2 - \frac{1}{4}a_{M(VO_2)}^2}}$ = 4.5498 \AA. From these values, the calculated in-plane strains are ${\epsilon_x}$ = 0.0171 and ${\epsilon_y}$ = 0.0097. Now substituting the values of ${\epsilon_x}$, ${\epsilon_y}$ and v = 0.25${^[}$\cite{aetukuri2013control}${^]}$ to eq. (8), the compressive out of plane strain is calculated to be 0.8962\%. This results an out of plane separation (${d_{(21\bar{1})_M})}$ of 2.4073 {\AA} in the strained films at monoclinic phase. From this, the out of plane XRD peak position for ${(21\bar{1})_M}$ plane can be calculated as 2${\theta}$ = 37.33 ° which is nearly equal to the experimentally observed out-of-plane XRD peak position of $\sim$37.4 ° at room temperature (Figure 3e) .
\begin{figure}[!htb]
	\centering
	\includegraphics[width=1\textwidth]{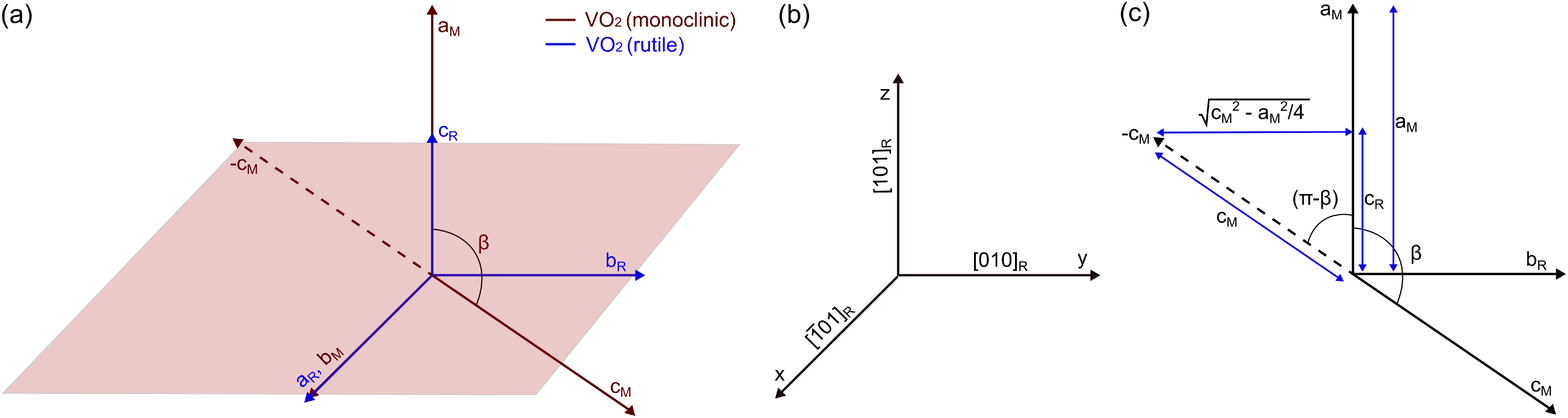}
	\caption{(a) Inter-relationship between VO$_2$ (rutile) and VO$_2$ (monoclinic) crystal co-ordinate system where a$_M$ $\approx$ 2c$_R$, b$_M$ $\approx$ a$_R$, c$_M$ $\approx$ b$_R$ - c$_R$. Plane shown in this figure is (101)$_R$ which is equivalent to (21${\bar{1}})_M$. (b) x - y plane represents the in-plane (101)$_R$ and z axis corresponds to the out of plane direction. (c) The geometrical representation for the in-plane component of (21${\bar{1})_M}$ plane on (101)$_R$ plane along [010] direction.}
\end{figure}
\nocite{*}
\bibliography{referanceVO2Work_supple}